\documentclass[a4paper,twocolumn,showpacs,nofootinbib,floatfix,superscriptaddress]{revtex4-1}
\usepackage{graphicx}
\usepackage{amsfonts}
\usepackage{amsmath}
\usepackage{hyperref}
\usepackage{float}

\renewcommand{\vec}[1]{\mathbf{#1}}
\def\be{\begin{equation}}
\def\ee{\end{equation}}
\def\bea{\begin{eqnarray}}
\def\eea{\end{eqnarray}}

\begin{document}

\title{Investigation of shock waves in the relativistic Riemann problem:
A comparison of viscous fluid dynamics to kinetic theory}

\author{I.\ Bouras}
\affiliation{Institut f\"ur Theoretische Physik,
Johann Wolfgang Goethe-Universit\"at,
Max-von-Laue-Strasse\ 1, D-60438 Frankfurt am Main, Germany}

\author{E.\ Moln\'ar}
\affiliation{Frankfurt Institute for Advanced Studies,
Ruth-Moufang-Strasse\ 1, D-60438 Frankfurt am Main, Germany}
\affiliation{KFKI, Research Institute of Particle and Nuclear Physics,
H-1525 Budapest, P.O.Box 49, Hungary}

\author{H.\ Niemi}
\affiliation{Frankfurt Institute for Advanced Studies,
Ruth-Moufang-Strasse\ 1, D-60438 Frankfurt am Main, Germany}

\author{Z.\ Xu}
\affiliation{Institut f\"ur Theoretische Physik,
Johann Wolfgang Goethe-Universit\"at,
Max-von-Laue-Strasse\ 1, D-60438 Frankfurt am Main, Germany}
\affiliation{Frankfurt Institute for Advanced Studies,
Ruth-Moufang-Strasse\ 1, D-60438 Frankfurt am Main, Germany}

\author{A.\ El}
\author{O.\ Fochler}
\author{C.\ Greiner}
\affiliation{Institut f\"ur Theoretische Physik,
Johann Wolfgang Goethe-Universit\"at,
Max-von-Laue-Strasse\ 1, D-60438 Frankfurt am Main, Germany}

\author{D.H.\ Rischke}
\affiliation{Institut f\"ur Theoretische Physik,
Johann Wolfgang Goethe-Universit\"at,
Max-von-Laue-Strasse\ 1, D-60438 Frankfurt am Main, Germany}
\affiliation{Frankfurt Institute for Advanced Studies,
Ruth-Moufang-Strasse\ 1, D-60438 Frankfurt am Main, Germany}

\begin{abstract}
We solve the relativistic Riemann problem in viscous matter
using the relativistic Boltzmann equation and the relativistic
causal dissipative fluid-dynamical approach 
of Israel and Stewart.
Comparisons between these two approaches 
clarify and point out the
regime of validity of second-order fluid dynamics in
relativistic shock phenomena.
The transition from ideal to viscous shocks is demonstrated 
by varying
the shear viscosity to entropy density ratio $\eta/s$.
We also find that a good agreement between these 
two approaches requires
a Knudsen number $Kn < 1/2$.

\end{abstract}


\date{\today}

\maketitle
\section{Introduction}

One of the strongest arguments for studying a fully relativistic
formulation of fluid dynamics is the large value of the elliptic flow
coefficient $v_2$ measured in ultrarelativistic heavy-ion
collisions at the Relativistic Heavy Ion Collider (RHIC) at 
Brookhaven National Laboratory (BNL) \cite{Adler:2003kt}.
These measurements indicate that a new phase of matter, the 
quark-gluon plasma, is created, which behaves like an
almost perfect fluid. 
An indication for this is the good agreement between
experimental data and theoretical calculations using ideal
fluid dynamics, where the viscous transport coefficients
vanish; see, e.g. Refs.,~\cite{Kolb:2003dz,Huovinen:2003fa,Huovinen:2006jp}.

In nature, the viscous transport coefficients cannot vanish but must have lower 
bounds \cite{Danielewicz:1984ww,Kovtun:2004de,Csernai:2006zz}.
Small viscous corrections are always needed in fluid dynamics to make 
the flow laminar and stable. It was later confirmed by
calculations within viscous fluid dynamics \cite{Luzum:2008cw}
and microscopic transport theory \cite{Xu:2007jv} that the shear 
viscosity has to be sufficiently small in order to keep the agreement 
with the $v_2$ data. These calculations suggest that the shear viscosity to 
entropy density ratio $\eta/s < 0.4$.

In heavy-ion collisions at ultrarelativistic energies the system expands 
very fast and gradients in the matter are very large.
It is still an open question to what extent fluid dynamics is applicable in 
describing the dynamics of such a system. Fluid dynamics is an effective 
theory that describes the macroscopic evolution of the system close to 
thermal equilibrium. Its applicability requires that either the viscosity 
or the gradients are small, or both. On the other hand, microscopic
transport models can be used for systems which are also strongly out of
thermal equilibrium. Therefore, a
comparison between the microscopic approach and fluid dynamics can provide
the limits and accuracy of the fluid-dynamical description. 

In this work we will compare solutions of the relativistic Boltzmann
equation with those of second-order fluid dynamics
as derived by Israel and Stewart (IS) 
\cite{Stewart:1972hg,Stewart:1977,Israel:1979wp}.
Such a comparison has been previously presented for the Bjorken
scaling solution \cite{Bjorken:1982qr} in Ref.~\cite{Huovinen:2008te}.
The scaling solution provides a simple test case with (initially) 
arbitrarily large expansion rate. It was concluded that for a 
good agreement between the fluid-dynamical
and kinetic calculations a Knudsen number $Kn<1/2$ is required. 
The Knudsen number is the ratio of the mean-free path of the particles
and the length scale of the variation of macroscopic fields (such as the energy
density). 
For the same scaling solution, an extension of the IS theory was 
studied and compared to the kinetic solutions in Ref.~\cite{El:2009vj}.

The Bjorken scaling solution is, however, restrictive in the sense 
that there are no pressure gradients in the local rest frame (LRF) of
matter. In this paper we will complement these earlier works by studying 
the relativistic Riemann problem.
This was already investigated in Refs.~\cite{Bouras:2009nn,Bouras:2009zz}, 
and we continue this line of study with more extensive and detailed comparisons 
between the fluid-dynamical and transport approaches. 
The existing analytic solution in the perfect-fluid limit, i.e., 
for $\eta = 0$, 
makes it possible to check the numerical convergence of the approaches. 
To our best knowledge this type of study was never accomplished for relativistic 
shock phenomena.

The present work is yet another step toward understanding how
highly-energetic jets created in the initial stage of an 
ultrarelativistic heavy-ion collision interact with the hot and dense 
medium. If the latter is a strongly interacting fluid, one
expects the creation of shock waves which resemble Mach cones. However,
if matter is only weakly interacting, these shock waves
should be smeared out and a clean Mach-cone signal cannot be observed
\cite{Bouras:2010nt}.

The paper is organized as follows: In Sec.~\ref{sec:kinetic_to_IS}
the dissipative fluid-dynamical theory of Israel and Stewart is introduced.
This is followed in Sec.~\ref{sec:numerics} by a short introduction to the 
numerical methods we use. 
In Sec.~\ref{sec:tests} the relativistic Riemann problem and its 
analytic solution in the perfect-fluid limit are presented. 
We demonstrate that numerical calculations using both approaches reproduce the 
analytic solutions very accurately.
In Sec.~\ref{sec:viscous_riemann} we show comparisons with non-zero
viscosities and demonstrate how the agreement between the two approaches 
deteriorates with increasing viscosity. 
These results are then analyzed in terms of a relevant Knudsen number 
in order to estimate the applicability of the IS theory.
As in the ideal case, we have found a scaling behavior of the
relativistic Riemann problem
with non-zero viscosity. This is discussed in Sec.~\ref{sec:scaling_behaviour}.
Finally, conclusions are given in Sec.~\ref{sec:conclusions}.

Our units are $\hbar = c = k = 1$; the metric is
$g^{\mu \nu} = \textrm{diag}(1,-1,-1,-1)$.

\section{From kinetic theory to fluid dynamics}
\label{sec:kinetic_to_IS}

\subsection{Definitions}

In relativistic kinetic theory of simple gases matter is
characterized by the invariant single-particle distribution function
$f(x,p)$.  The space-time evolution of $f(x,p)$ caused by particle
motion and collisions is given by the relativistic Boltzmann
transport equation \cite{deGroot,Cercignani_book},
\be\label{BTE}
p^{\mu} \partial_{\mu} f(x,p) = C \left[f(x,p) \right] \, ,
\ee
where $C \left[f(x,p) \right]$ is the collision integral and
$p^{\mu} = (E,\vec{p})$ is the particle four-momentum. We assume
that there are no external forces.

Macroscopic quantities can be obtained from the moments of the
distribution function.  The first moment is the particle
four-flow,
\be\label{kinetic:N_mu}
N^{\mu} \equiv \int d\tilde{p} \, p^{\mu} f(x,p) \, ,
\ee
where $d\tilde{p} \equiv g\,d^3 p/[(2 \pi)^3 E]$ and $g$ is
the degeneracy factor counting internal degrees of freedom.
The second moment defines the energy momentum tensor,
\be\label{kinetic:T_mu_nu}
T^{\mu \nu} \equiv \int d\tilde{p} \, p^{\mu} p^{\nu} f(x,p) \, .
\ee
Similarly one can define higher moments of the distribution function, such as 
the third moment, 
\be\label{kinetic:F_mu_nu_lam}
F^{\mu \nu \lambda} \equiv \int d\tilde{p}\, 
p^{\mu} p^{\nu} p^{\lambda} f(x,p) \, .
\ee
The entropy four-current is defined as
\be\label{kinetic:S_mu}
S^{\mu} \equiv \int d\tilde{p} \, p^{\mu}
f(x,p) \left[ 1 - \ln f(x,p) \right] \, .
\ee

The assumption of the conservation of particle number or charge and
energy-momentum conservation in individual collisions leads to
\bea\label{cons_N_mu}
\partial_{\mu} N^{\mu}  
&\equiv& \int d\tilde{p}\, p^{\mu} \partial_{\mu} f
= \int d\tilde{p} \, C = 0 \, , \\ 
\label{cons_T_mu_nu}
\partial_{\nu} T^{\mu\nu} 
&\equiv& \int d\tilde{p}\, p^{\mu} p^{\nu} \partial_{\nu} f
= \int d\tilde{p} \, p^{\mu} C = 0 \, .
\eea
These are the conservation equations of relativistic
fluid dynamics \cite{Taub:1948zz}.
A similar equation resulting from the Boltzmann equation for the third moment 
gives the balance of fluxes,
\be
\label{cons_F_mu_nu_lam}
\partial_\lambda F^{\mu \nu \lambda} 
\equiv \int d\tilde{p} \, p^{\mu} p^{\nu} p^{\lambda} \partial_\lambda f
= \int d\tilde{p} \, p^{\mu} p^{\nu} C \, .
\ee
This will be discussed in the next section.
The Boltzmann H-theorem implies that the entropy production
is positive and vanishes in equilibrium 
\be
\partial_{\mu} S^{\mu} \ge 0 \, .
\ee

The particle four-flow and energy-momentum tensor 
can be decomposed with respect to an arbitrary time-like
four-vector $u^\mu$, normalized as $u^\mu u_\mu = 1$. 
This is chosen in such a way that it can be interpreted as the collective 
four-velocity of the matter.
The frame where $u^\mu = (1, 0, 0, 0)$ is called the local rest frame.
With the help of the transverse projection operator
$\Delta^{\mu \nu} = g^{\mu\nu} - u^\mu  u^\nu$, the 
most general decomposition can be written as
\bea
\label{N_mu}
N^{\mu} &=& n u^{\mu} + V^{\mu}  \, , \\
\label{T_munu}
T^{\mu \nu} &=& e u^{\mu} u^{\nu} - P\Delta^{\mu \nu}
+ W^{\mu}u^{\nu} + W^{\nu}u^{\mu} + \pi^{\mu \nu} \, , \qquad
\eea
where $n \equiv N^{\mu}u_{\mu}$ is the LRF particle density
and $e \equiv u_{\mu} T^{\mu \nu} u_{\nu}$ is the LRF energy density.
The spatial trace of the energy-momentum tensor,
$P = -\frac{1}{3}\Delta_{\mu \nu} T^{\mu \nu}$ denotes the isotropic pressure.
This is the sum of equilibrium 
pressure and bulk viscous pressure, $P = p(e,n) + \Pi$.
The flow of particles in the LRF is 
\be
V^{\mu} = \Delta^{\mu}_{\nu} N^{\nu}\;,
\ee
while the flow of energy-momentum in the LRF is
\be
W^{\mu} = \Delta^{\mu \alpha} T_{\alpha \beta} u^{\beta}\;.
\ee
The heat flow $q^\mu$ is defined as
\be
q^\mu = W^\mu - h\, V^\mu\;,
\ee
where $h = (e+p)/n$ is the enthalpy per particle.

The shear-stress tensor, $\pi^{\mu \nu} = T^{\langle \mu \nu \rangle}$, where
\be
T^{\langle \mu \nu \rangle} 
\equiv \left[\frac{1}{2}\left( \Delta^{\mu}_{\alpha} \Delta^{\nu}_{\beta} +
\Delta^{\nu}_{\alpha} \Delta^{\mu}_{\beta} \right) - \frac{1}{3}
\Delta^{\mu \nu} \Delta_{\alpha \beta} \right] T^{\alpha \beta} \, ,
\ee
is that part of $T^{\mu \nu}$, that is symmetric, traceless, and
orthogonal to the flow velocity.

In equilibrium all dissipative terms vanish, that is, 
$\Pi=V^{\mu}=W^{\mu}=\pi^{\mu \nu}=0$, therefore the particle 
four-current and energy-momentum tensor take a simpler form, 
\bea \label{N_mu_eq}
N^{\mu}_0 &=& n_0 u^{\mu}   \, , \\ \label{T_mu_nu_eq}
T^{\mu \nu}_0 &=& e_0 u^{\mu} u^{\nu} - p_0\Delta^{\mu \nu} \, ,
\eea
where the subscript ``0'' indicates local thermodynamical equilibrium.
In local equilibrium, the particle four-current $N^{\mu}_0$,
the energy-momentum tensor $T^{\mu \nu}_0$, and the entropy
four-current, $S^{\mu}_0$, are uniquely
defined for any time-like four-velocity $u^{\mu}$.
The LRF frame particle density and energy density are given by
$n_0 \equiv N^{\mu}_0 u_{\mu}$ and 
$e_0 \equiv u_{\mu} T^{\mu \nu}_0 u_{\nu}$, respectively.

For a system that deviates from local thermodynamical equilibrium,
the definition of the flow velocity may follow Eckart
\cite{Eckart:1940te} or 
Landau and Lifshitz \cite{Landau_book}.
Using Eckart's definition of the flow velocity,
\be\label{eq_hydro_kinetic:eckart_velocity}
u^{\mu}_{E} = \frac{ N^{\mu} }{ \sqrt{ N^{\nu} N_{\nu} } } \, ,
\ee
the LRF flow of particles vanishes, $V^{\mu} = 0$, while
the flow of energy is given by the heat flow,
$W^{\mu} = q^{\mu}$.

In Landau's frame the flow of matter is tied to the flow
of energy,
\be\label{eq_hydro_kinetic:landau_velocity}
u^{\mu}_{L} = \frac{ T^{\mu \nu} u_{\nu} }
{ \sqrt{ u_{\alpha} T^{\beta \alpha} T_{\beta \gamma} u^{\gamma}} } \, ;
\ee
hence the flow of energy-momentum vanishes, $W^{\mu} = 0$.

For processes close to equilibrium, the above definitions
and decompositions are related to each other \cite{Israel:1976tn}.
Therefore up to second order in deviations from local equilibrium, 
that is, $\delta u^{\mu} \sim q^{\mu}/e \ll 1$, we get
\bea
u^{\mu}_L = u^{\mu}_E + \frac{q^{\mu}}{e + p}\, .
\eea
This means that the non-equilibrium part of the particle
four-flow in Landau's frame is related to the heat flow
in Eckart's frame, that is, $V^{\mu} = -q^{\mu}/h$.

\subsection{Fluid dynamics as an approximation to kinetic theory}

Fluid dynamics is an effective theory for the slow,
long-wavelength dynamics of a given system. For systems with 
well-defined quasi-particles, fluid dynamics
can be derived in terms of a power series in the Knudsen number
\be
Kn = \frac{\lambda_{\rm mfp}}{L} \, ,
\ee
where $\lambda_{\rm mfp}$ is the mean-free path of the particles
and $L$ is a macroscopic length scale over which macroscopic fields
such as energy density, particle density, or temperature vary.
Fluid dynamics as an effective theory can be systematically improved by 
successively including higher-order terms in $Kn$ \cite{El:2009vj}. 

To zeroth order in $Kn$ we obtain an effective theory that does
not contain any powers of $Kn$, corresponding to the limit $Kn \rightarrow
0$, that is, the (unphysical) limit where $\lambda_{\rm mfp} \rightarrow 0$.
This corresponds to infinite scattering rates, and thus the system
instantaneously assumes local thermodynamical equilibrium. This is
the perfect-fluid limit.
To first order in $Kn$, we obtain the relativistic generalization of
Navier-Stokes theory. This effective theory is plagued by
instabilities and acausalities
\cite{Hiscock:1983zz,Hiscock:1985zz,Pu:2009fj}.
These problems can be circumvented by including terms of second
order in $Kn$, such as in the fluid-dynamical theory of Israel and
Stewart used in this work. The fluid-dynamical 
limit can be derived for any kind of
system, that is, its applicability is not restricted to dilute gases, as is
the case for the Boltzmann equation. However, since it is an
expansion around the perfect-fluid limit, we
expect its validity to be restricted to dynamics close to local
thermodynamical equilibrium.

The coefficients of fluid dynamics as an effective theory can be 
computed by matching to an underlying microscopic theory. 
In our case, this is the kinetic
theory of ultrarelativistic Boltzmann particles, described by the
Boltzmann equation with elastic binary collisions.
Note, however, that the matching procedure is not unique \cite{Denicol:2010xn}.

In our case the matching procedure is as follows.
We expand the single-particle distribution function 
around local thermodynamical equilibrium,
$f(x,p) = f_0(x,p) + \delta f$,
where $\delta f$ measures the deviation from
the equilibrium distribution function 
\be \label{eq_distribution}
f_{0}(x,p) \equiv (e^{\beta p^{\mu} u_{\mu}}/\lambda + a)^{-1} \, .
\ee
Here $a = -1 (+1)$ corresponds to bosons (fermions), and $a=0$
corresponds to Boltzmann particles.
The inverse temperature is given by $\beta(x) = 1/T(x)$ and
$\lambda(x) = e^{\beta(x) \mu(x)}$ is the fugacity.
The validity of the expansion around $f_0$ requires that
$\phi = \delta f/f_0 \ll 1$.
In order to match to fluid dynamics as an expansion
in powers of $Kn$, one possibility is to assume that 
$\phi$ is a series in powers of $Kn$.
This approach was pioneered by Hilbert, and by Chapman and Enskog 
\cite{Cercignani_book}.

Another method was proposed by Grad \cite{Grad} and was generalized
to relativistic systems by Israel and Stewart 
\cite{Stewart:1972hg,Stewart:1977,Israel:1979wp}.
In this approach,
\be
\label{eq:f_expansion}
\phi(x,p) = \epsilon + \epsilon_{\mu} p^{\mu}
+ \epsilon_{\mu \nu} p^{\mu} p^{\nu}\, ,
\ee
where the coefficients $\epsilon$, $\epsilon_{\mu}$, and $\epsilon_{\mu
\nu}$ are the expansion parameters and therefore $\sim O(Kn)$.

For each non-equilibrium state given by $f$ the corresponding 
equilibrium state $f_0$ is defined by the Landau matching 
conditions \cite{Israel:1979wp}, such that  $n_0=n$ and $e_0=e$.
These conditions together with the Gibbs equation ensure that the equilibrium 
part of the pressure is given by the equation of state
$p_0(e,n) = p_0(e_0,n_0)$.

The coefficients of the expansion, $\epsilon, \epsilon_{\mu}$, 
and $\epsilon_{\mu \nu}$, are determined using 
Eq.~\eqref{eq:f_expansion} in Eqs.~\eqref{kinetic:N_mu} and 
\eqref{kinetic:T_mu_nu}.
Comparing the outcome with Eqs.~\eqref{N_mu} and \eqref{T_munu}, 
one finds that the parameters of
the expansion are proportional to the dissipative quantities 
$\Pi, V^{\mu}, q^{\mu}$, and $\pi^{\mu \nu}$.
Thus the deviation from equilibrium is proportional to the 
ratio of dissipative quantities to local equilibrium quantities, that is, 
$\phi \sim \Pi/e,q^\mu/e,\pi^{\mu\nu}/e$.
However, if the dissipative quantities are close to the 
Navier-Stokes values (see later), then these ratios are proportional 
to the local Knudsen number.

The macroscopic equations for the evolution of dissipative quantities 
can be obtained from the third moment of the single-particle 
distribution function \cite{Israel:1979wp}.
Here we recall the result of this laborious calculation by 
Israel and Stewart as presented by Huovinen and Moln\'ar \cite{Huovinen:2008te}.
In this paper we consider only massless gases in which case bulk viscosity
vanishes. Then the IS equations for the heat flow and the shear-stress tensor are, 
\bea \label{q_IS_full}
Dq^\mu & = & \frac{1}{\tau_q}\left( q^\mu_{NS} - q^\mu \right)
- \omega^{\mu\lambda} q_\lambda \\ \nonumber
& & - u^\mu q_\nu D u^\nu - \frac{1}{2} q^\mu\left(\nabla_\lambda u^\lambda
+ D \ln \frac{\beta_1}{T}\right)
 \\ \nonumber
& &
+ \frac{\alpha_1}{\beta_1}(\partial_\lambda \pi^{\lambda\mu}
+ u^\mu \pi^{\lambda\nu} \partial_\lambda u_\nu)
-\frac{a_1}{\beta_1} \pi^{\lambda\mu} D u_\lambda \, ,\\ \label{pi_IS_full}
D\pi^{\mu\nu} &=& \frac{1}{\tau_\pi}\left(\pi^{\mu\nu}_{NS} - \pi^{\mu\nu} \right)
- 2 \pi_\lambda^{\ \langle\mu}\omega^{\nu\rangle\lambda}  \\ \nonumber
& & - (\pi^{\lambda\mu}u^\nu+\pi^{\lambda\nu}u^\mu)D u_\lambda \\ \nonumber
& & -\frac{1}{2}\pi^{\mu\nu}\left(\nabla_\lambda u^\lambda
+ D \ln \frac{\beta_2}{T}\right) \\ \nonumber
& & -\frac{\alpha_1}{\beta_2} \nabla^{\langle\mu} q^{\nu\rangle}
+ \frac{a_1^\prime}{\beta_2} q^{\langle\mu} D u^{\nu\rangle} \, ,
\eea
where the proper-time derivative is denoted by
$D \equiv u^{\mu} \partial_{\mu}$,
the gradient operator is
$\nabla^{\mu} = \Delta^{\mu \nu} \partial_{\nu}$,
and the vorticity tensor is
$\omega^{\mu \nu}= \frac{1}{2}\Delta^{\mu \alpha}
\Delta^{\beta \nu} \left(\partial_{\beta}u_{\alpha}
- \partial_{\alpha} u_{\beta} \right)$.
The coefficients $\alpha_1, \beta_1, \beta_2$, and $a_1$ are thermodynamic
functions, and $a'_1 = \left(\partial(\beta \alpha_1)/\partial \beta\right)_{\mu/T} - a_1$.
These depend whether we choose the Landau or Eckart frame.
The Navier-Stokes values for the heat flow and shear-stress tensor are
\bea\label{q_NS}
q^\mu_{NS} &\equiv& - \kappa_q\frac{n T^2}{e + p}
\nabla^\mu \left(\frac{\mu}{T}\right) \, , \\
\label{pi_NS}
\pi^{\mu\nu}_{NS} &\equiv& 2\eta\nabla^{\langle\mu}u^{\nu\rangle} \, ,
\eea
where $\kappa_q$ is the heat conductivity coefficient and
$\eta$ is the shear viscosity coefficient.
The relaxation times of heat conductivity and shear viscosity are proportional
to the heat conductivity and shear viscosity coefficient, respectively, 
that is, $\tau_q = \kappa_q T \beta_1$ and $\tau_\pi = 2\eta \beta_2$. 

The microscopic time scales in the IS equations are given by the
relaxation times of dissipative quantities $\tau_\pi$ and $\tau_q$, 
which are of the order of the mean-free path between collisions.
The relevant macroscopic scales can be estimated from the gradients
of the primary fluid-dynamical variables.
For example, these can be given in terms of the expansion rate $L_{\theta} = 1/\theta$, 
in terms of the energy density gradient $L_{e}^{-1} = \sqrt{\nabla^\mu e \nabla_\mu e}/e$, or
in other ways.

If the Knudsen number is sufficiently small, then at late times 
$t > \tau_\pi, \tau_q$,
heat flow and shear viscosity will approach their Navier-Stokes values, that is, 
$q^\mu \sim q_{NS}^\mu$ and $\pi^{\mu\nu} \sim \pi_{NS}^{\mu\nu}$. When this
happens, the dissipative quantities can be estimated to be of order
$1$ in the Knudsen number, and Eqs.~\eqref{q_IS_full} and
\eqref{pi_IS_full} include contributions 
up to second order in $Kn$.

In the limit $\tau_q, \tau_\pi \rightarrow 0$, with $\eta$ and $\kappa_q$
constant, the IS equations reduce to the Navier-Stokes equations.
Furthermore, in the limit when all dissipative quantities approach zero,
the IS equations reduce to the perfect-fluid equations.

\subsection{The Israel-Stewart equations for (1+1)--dimensional expansion}
\label{subsec:IS_1plus1}

For the sake of simplicity, we assume an ultrarelativistic 
massless Boltzmann gas with conserved particle number.
In this case, the bulk viscosity vanishes, and the equation of state 
is simply $e = c^{-2}_s p$, where the speed of sound is $c_s = \sqrt{1/3}$.
For gluons, the energy density as a function of temperature is
$e = 3 n T$, where $n = \lambda\, g T^3/\pi^2$ is the number density,
with $g = 16$ being the gluon number of degrees of freedom.
The entropy density is given by
$s = (4 - \ln \lambda) n$.

In the following we choose the Landau frame.
We shall briefly discuss and write the IS equations
in (1+1)--dimensional Cartesian coordinates.
We assume that the system is homogeneous in the transverse directions,
$x$ and $y$, and evolves along the longitudinal direction $z$ such that
the velocities as well as the derivatives in both transverse directions
vanish identically.
Thus the four-velocity is $u^{\mu} = \gamma_z(1,0,0,v_z)$ where
$\gamma_z = (1 - v^2_z)^{-1/2}$, while
the four-derivative is $\partial_\mu = (\partial_t, 0, 0, \partial_z)$.
The following four-vector and tensor components vanish:
$N^{x} = N^{y} = 0$ and $T^{0x}=T^{0y}=T^{xy}=T^{xz}=T^{yz}=0$.
This also implies that the heat-flow components
$q^x = q^y = 0$ and shear-stress tensor components
$\pi^{0x}=\pi^{0y}=\pi^{xy}=\pi^{xz}=\pi^{yz}=0$
vanish identically.

Using the orthogonality of the heat-flow four-vector
we obtain that $q^0 = q^z v_z$.
We may also define the magnitude of the heat-flow four-vector
by $q = \sqrt{-q^{\mu} q_{\mu}}$, thus $q^z = \gamma_z q$.
Similarly, using the orthogonality property of the
shear-stress tensor we get $\pi^{00} = \pi^{0z} v_z$ and
$\pi^{0z} = \pi^{zz} v_z$.
To satisfy the tracelessness condition we may choose
$\pi^{xx} = \pi^{yy} = -\pi/2$ and $\pi^{zz} = \gamma^2_z \pi$.

Therefore, the non-vanishing components of the particle four-current
and energy-momentum tensor are
\bea
N^{0} &\equiv& n\gamma_z - \frac{q^z v_z}{h}\, , \\
N^{z} &\equiv& N^{0} v_z - \frac{q^z}{\gamma^2_z h}\, , \\
T^{00} &\equiv& (e + \mathcal{P}_z)\gamma^2_z - \mathcal{P}_z \, , \\
T^{0z} &\equiv& v_z (T^{00} + \mathcal{P}_z) \, , \\
%
T^{xx} &\equiv& p - \frac{\pi}{2} = T^{yy} \, , \\
T^{zz} &\equiv& v_z T^{0z} + \mathcal{P}_z \, ,
\eea
where the LRF effective pressure is,
\bea
\mathcal{P}_z &=& p(e,n) + \pi \, .
\eea
The LRF particle and energy densities expressed through
the laboratory frame quantities and the velocity are
\bea
n &=& N^0 \left[(1 - v^2_z)^{-1/2} - \frac{q^z v_z}{e + p} \right]^{-1} \, , \\
e &=& T^{00}  - v_z T^{0z} = T^{00} - \frac{(T^{0z})^2}{T^{00} + \mathcal{P}_z}\, ,\\
v_z &=& \frac{T^{0z}}{T^{00} + \mathcal{P}_z} \, .
\eea
The conservation equations are
\bea
\partial_t N^{0} + \partial_z (v_z N^{0})
&=& \partial_z \left[ \frac{q^z \, n}{\gamma^2 (e+p)}\right] \, , \label{cons_N0}\\
\partial_t T^{00} + \partial_z (v_z T^{00})
&=& -\partial_z(v_z \mathcal{P}_z) \label{cons_T00}\, , \\
\partial_t T^{0z} + \partial_z (v_z T^{0z})
&=& -\partial_z \mathcal{P}_z \label{cons_T0z}\, .
\eea
The relaxation equations for the heat conductivity are calculated from
Eqs.~\eqref{q_IS_full} and \eqref{pi_IS_full}. In the (1+1)--dimensional case
the terms containing the vorticity vanish; therefore the relaxation equations
can be written formally as
\bea \label{relaxationEq_heatAndShear}
D q^z & = & \frac{1}{\tau_q}\left(q^z_{NS} - q^z\right)
- I^z_{q1} - I^z_{q2} - I^z_{q3}\, , \\
D\pi &=& \frac{1}{\tau_\pi}\left(\pi_{NS} - \pi \right)
- I_{\pi1} - I_{\pi2} - I_{\pi 3}\, ,
\eea
where the Navier-Stokes values for the heat conductivity and shear stress are,
\bea
\pi_{NS} &=& - \frac{4}{3} \left( \frac{\eta}{s} \right) s \, \theta_z \, , \\
q^z_{NS} &=&  \left( \frac{\kappa_q T}{s} \right) \frac{(T s) n}{e + p} \,
\gamma^2_z \left( v_z \frac{\partial_t \lambda}{\lambda}
+ \frac{\partial_z \lambda}{\lambda} \right) \, .
\eea
The expansion rate is denoted by
$\theta_z = \partial_t \gamma_z + \partial_z(\gamma_z v_z)$.
In the ultrarelativistic limit,  $\alpha_1 = -1/(4p)$, $\beta_1 = 5/(4p)$, $\beta_2 = 3/(4p)$,
$a_1 = 0$, and $a'_1 = 5 \alpha_1$. 
The terms in the relaxation equations are given explicitly as
\bea
I^z_{q1} &=& \frac{1}{2}q^z
\left(\theta_z + D \ln \frac{\beta_1}{T} \right) \, ,\\
I^z_{q2} &=& - q^z v_z \gamma^3_z \left(\partial_t v_z
+ v_z \partial_z v_z \right) \, ,\\
I^z_{q3} &=& \frac{1}{5} \left[ \gamma^2_z \left( v_z \partial_t \pi + \partial_z \pi\right)
+  \gamma_z \pi \left(v_z \theta_z + \gamma_z \partial_t v_z \right)\right] \, , \qquad
\eea
and
\bea
I_{\pi 1} &=& \frac{1}{2}\pi
\left(\theta_z + D \ln \frac{\beta_2}{T} \right) \, ,\\
I_{\pi 2} &=& \frac{10}{9} (q^z\gamma^2_z)\left(\partial_t v_z
+ v_z \partial_z v_z \right) \, ,\\
I_{\pi 3} &=& \frac{2}{9} \left(v_z \partial_t q^z + \partial_z q^z
- \frac{q^z v_z}{\gamma_z} \theta_z \right) \, .
\eea

The terms $I^z_{q3}$, $I_{\pi 2}$, and $I_{\pi 3}$ represent a
coupling between the heat-flow four-vector and shear-stress tensor.

In this work the term $I^z_{q3}$ is neglected in most cases 
unless otherwise stated. 
The reason is that the agreement with kinetic theory is better without it, 
but results with and without this coupling term will be shown when we discuss 
viscous solutions to the Riemann problem.

\section{Numerical methods} 
\label{sec:numerics}
\subsection{The transport model: BAMPS}
\label{sec:BAMPS}

The microscopic transport model we use is the
Boltzmann approach of multiparton scatterings (BAMPS)
\cite{Xu:2004mz,Xu:2007aa}.
This numerical method solves the Boltzmann equation \eqref{BTE}
for on-shell particles based on the stochastic interpretation
of transition rates. In this study we consider only binary collisions
with an isotropic cross section, that is, a cross section with an isotropic
distribution of the collision angle.

Simulations of the space-time evolution of particles are performed
in a static box. Because we consider shock-wave propagation in one 
dimension as described in the next section, we choose the $z$-axis as
the direction along which shock waves propagate. The $x$- and $y$-axes
span the transverse plane. 

The whole box is divided into spatial cells with a volume
$V_{\rm cell} = \Delta x \, \Delta y \, \Delta z$. Collisions of
particles in the same cell are simulated by Monte Carlo technique according
to the individual collision probability within a time step $\Delta t$,
\begin{equation}\label{eq_BAMPS_collProb}
P_{ \rm 22} = v_{ \rm rel} \frac{ \sigma }
{ N_{\rm test} } \frac{\Delta t}{ V_{\rm cell} } \, ,
\end{equation}
where $\sigma$ is the total cross section, and
$v_{ \rm rel} = (p_1 + p_2)^2/(2 E_1 E_2)$ denotes the relative
velocity of the two incoming particles with four momenta $p_1, p_2$.
In order to reduce statistical fluctuations in simulations and to
ensure an accurate solution of the Boltzmann equation \eqref{BTE}
a testparticle method \cite{Xu:2004mz} is introduced: The particle number
is artificially increased by multiplying it by the number of
testparticles per real particle, $N_{\rm test}$.
Thus, the collision probability has to be reduced by the same number to
keep the particle mean-free path independent of the value of $N_{\rm test}$.

If a collision occurs, momenta of colliding particles are changed according
to the isotropic distribution of the collision angle. Before and after
collisions particles propagate via free streaming. Collisions of particles
against box boundaries are realized as elastic collisions off a wall. 
We have to stop simulations before the shock front reaches the box
boundaries, otherwise it would get reflected.

The relationship between the shear viscosity $\eta$ and the total cross
section $\sigma$ is given by $\eta = 4 e / (15 R^{tr})$ \cite{Xu:2007ns},
where $R^{tr} = n \langle v_{\rm rel} \sigma^{tr} \rangle= 
2 n \langle v_{\rm rel} \sigma \rangle /3$ is the transport
collision rate in the case of isotropic scattering processes 
\cite{Huovinen:2008te}.  $\langle \rangle$ stands for the ensemble average
in the LRF. In IS theory, we obtain
\begin{equation}\label{eq_BAMPS_shearViscosity}
\eta = \frac{2}{5} e \lambda_{\rm mfp}
\end{equation}
for an ultrarelativistic massless gas, where
$\lambda_{\rm mfp} = 1/(n \langle v_{\rm rel} \sigma \rangle)$
is the particle mean-free path.
Note that $\langle v_{\rm rel} \sigma \rangle$ is inversely
proportional to $\lambda_{\rm mfp}$ and $\eta/s$. For given constant
$\lambda_{\rm mfp}$ or $\eta/s$, the average 
$\langle v_{\rm rel} \sigma \rangle$ in individual cells is
generally different, if there exists a spatial gradient of $n$ and/or $e$.
In these cases we assume that $\sigma$ does not depend on the momenta
of two incoming particles. This leads to 
$\langle v_{\rm rel} \sigma \rangle \approx \sigma$, because
$\langle v_{\rm rel} \rangle \approx 1$ in the LRF.

In the following, we shall also need the relationship 
between the heat conductivity 
and the cross section \cite{deGroot,Molnar:2007an} which,
in IS theory, is
\be
\kappa_q =  2/\sigma\, .
\ee

\subsection{The viscous hydro solver: vSHASTA}
\label{sec:vSHASTA}

In order to solve the IS equations of causal relativistic fluid dynamics
we use a version of the sharp and smooth transport algorithm (SHASTA) 
\cite{BorisBook_SHASTA}.
This numerical method is widely used in modeling relativistic heavy-ion 
collisions, and hence was extensively tested in the perfect-fluid 
approximation \cite{Schneider:1993gd,Rischke:1995ir}.
We apply SHASTA to solve both the conservation equations and the relaxation 
equations rearranged in conservation form and call this numerical method 
vSHASTA \cite{Molnar:2008fv}.

This algorithm requires the Courant-Friedrichs-Lewy (CFL) 
condition $\lambda_{\rm CFL}  = \Delta t/\Delta z \leq 0.5$,
where $\Delta t$ is the time step and $\Delta z$ is the cell size.
In all our numerical calculations we take $\lambda_{\rm CFL} = 0.4$.
In a first-order finite-difference approach this means that 
causal transport of matter covers only a distance 
$\lambda_{\rm CFL} \Delta z$,
while the remaining part of the matter is acausally diffused
over a distance $(1-\lambda_{\rm CFL})\Delta z$.
This purely numerical effect called prediffusion is 
partially removed by higher-order non-linear corrections in SHASTA.
The remaining low-order numerical diffusion represents
the so-called numerical viscosity of the algorithm.
In a strict sense numerical viscosity does not fully 
correspond to real physical viscosity, which is independent 
of the numerical method and resolution. 
However, its presence is inevitable and at the 
same time compulsory to keep the computations stable and 
to smooth out dispersion errors.
By increasing the numerical resolution this numerical diffusion can be 
always be reduced to smaller values than the physical one.  

We also mention another rather trivial numerical artifact which 
is present in the numerical solutions at early times.
The numerical solutions at early times do not represent the correct and 
accurate physical behavior, not even in the perfect-fluid limit.
However, this will change in time as the solution spreads over a
larger number of cells while the structures are resolved on a finer grid.
Therefore, the numerical solutions approach the correct solution 
only after some amount of time.

\section{Testing the perfect-fluid limit}
\label{sec:tests}
\subsection{The relativistic Riemann problem in the
perfect-fluid limit}
\label{sec:Riemann_problem}

In this section we introduce the relativistic Riemann problem 
in the case of perfect fluids. The matter is assumed
to be thermodynamically normal \cite{Rischke:1995mt} and, for
the sake of simplicity, to be
homogeneous in the transverse directions, so the problem
becomes ($1+1$)-dimensional.

In the Riemann problem we have matter in thermodynamical
equilibrium separated by a membrane at $z=0$. 
The pressures on the left $(z < 0)$ and right $(z\geq 0)$ sides of 
the membrane are $p_0$ and $p_4$, and the particle densities 
are $n_0$ and $n_4$, respectively. 
Here we only discuss a special case of the Riemann problem, 
called the shock-tube problem, when the velocities on 
both sides of the membrane are zero, that is, $v_0 = v_4 = 0$.

Removal of the membrane at time $t = 0$ leads to two propagating
waves. If $p_0 > p_4$, a shock wave is propagating to the
right with the velocity $v_{\rm shock}$. Simultaneously, the tail
of a rarefaction fan is propagating to the left with the speed
of sound $c_s$ into the matter with higher pressure. 
The region between these two waves includes a contact
discontinuity, propagating to the right with $v_3$, and
a shock plateau, which is bounded by the
contact discontinuity and the shock front.

\begin{figure}[th]
\includegraphics[width=8.6cm]{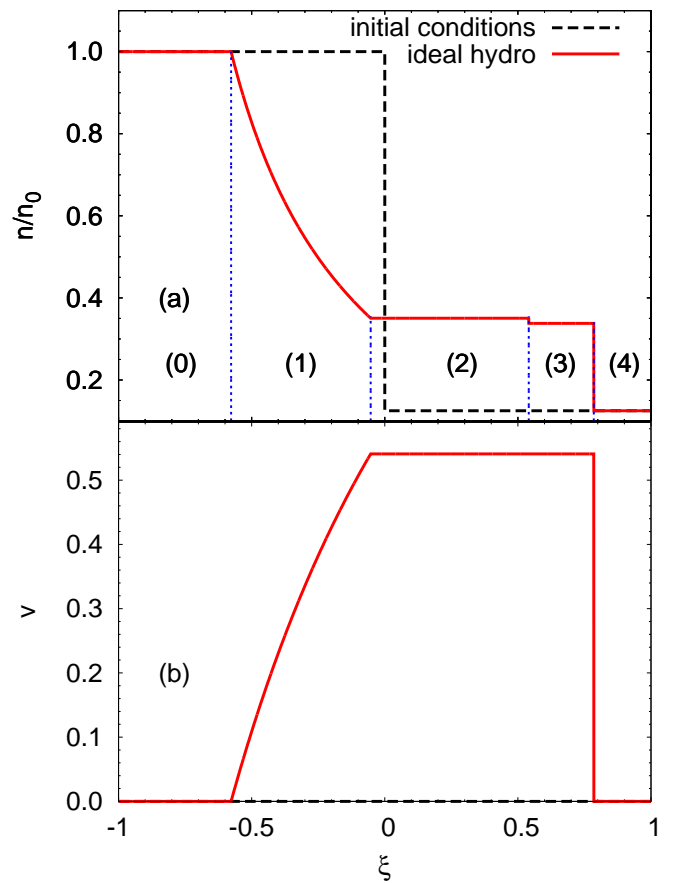}
\caption{(Color online) Analytic solution of the Riemann problem for 
a perfect fluid, (a) particle density and (b) velocity, as a function
of the similarity variable $\xi = z/t$.
The initial temperatures are $T_0 = 0.4$ GeV  
and $T_4 = 0.2$ GeV.} 
\label{fig:riemann_nv}
\end{figure}

Figure \ref{fig:riemann_nv} shows the analytic solution for the particle
density and velocity profile for an ultrarelativistic massless gas, $e = 3p$.
Here, regions $0$ and $4$ represent the undisturbed matter at rest,
$1$ is the rarefaction wave, $2$ denotes the constant
region between the tail of the rarefaction wave and the contact
discontinuity, while $3$ is the shock plateau.
The shock front is the discontinuity between regions $3$ and $4$.

%
%

The solution of the relativistic Riemann problem
is obtained by matching the pressure $p_2$ and velocity $v_2$ 
at the rarefaction tail to the pressure and velocity of 
the shock plateau $p_3$ and $v_3$, that is,  
$p_2 = p_3$ and $v_2 = v_3$; see Ref.\ \cite{Schneider:1993gd}
for more details.

The solution at the discontinuity is given by the
Rankine-Hugoniot-Taub relations \cite{Taub:1948zz} in the LRF
of the shock,
\begin{equation}
\begin{split}
n_3 \widehat{u}_3 &= n_4 \widehat{u}_4 \, , \\
(e_3 + p_3) \widehat{\gamma}_3 \widehat{u}_3 
& = (e_4 + p_4) \widehat{\gamma}_4 \widehat{u}_4 \, ,\\
(e_3 + p_3) \widehat{u}_3^{2} + p_3 
& = (e_4 + p_4) \widehat{u}_4^{2} + p_4 \, .
\end{split}
\end{equation}
Quantities with a ``$\;\widehat{}\;$'' are evaluated
in the LRF of the shock. 
Then, the velocities of the shock plateau 
and the shock front in the LRF of the undisturbed matter can
be expressed in terms of thermodynamic quantities before and after
the discontinuity,
\begin{align}
v_{\mathrm{plat}} &\equiv v_3 = \left [ \frac{(p_3 - p_4)(e_3 - e_4)}
{(e_4 + p_3)(e_3 + p_4)} \right]^{\frac{1}{2}} \label{eq_taub:v_3} \, ,\\
v_{\mathrm{shock}} &\equiv - \widehat{v_4}
= \left [ \frac{(p_4 - p_3)(e_3 + p_4)}{(e_4 - e_3)(e_4 + p_3)} \right]^{\frac{1}{2}} \, ,
\label{eq_taub:v_shock}
\end{align}
where $e_i = 3p_i$, $i=3,4$, for an ultrarelativistic gas of massless
particles.

The solution for the ideal shock-tube problem is self-similar in time, that is, 
the solution keeps the same shape at all times, $t>0$, without change.
This is best seen if we plot the solutions against the similarity variable, 
$\xi = z/t$, as is done in Fig.\ \ref{fig:riemann_nv}.

\subsection{Numerical convergence of BAMPS}
\label{sec:cell_dependence}

Before we employ BAMPS to solve the Riemann problem near the
perfect-fluid limit in the next subsection, we first show
the convergence of BAMPS  when the model parameters are varied.

In BAMPS the relevant parameters that control the numerical
accuracy are the cell size $\Delta z$, the time step $\Delta t$, and 
the test-particle number per real particle, $N_{\rm test}$. 
The cell sizes in the transverse
directions, $\Delta x$ and $\Delta y$, are not relevant, because the
system is assumed to be homogeneous in the transverse
plane. In addition, $\Delta t$ is always chosen to be smaller than $\Delta z$,
to avoid possible large local variations within one time step.
If one decreases $\Delta z$, one has to simultaneously increase $N_{\rm test}$ 
to ensure that each cell contains a sufficiently large number of
test particles. Thus, using BAMPS
the Boltzmann equation \eqref{BTE} will be exactly solved 
in the limit $\Delta z \rightarrow 0$ and $N_{\rm test} \rightarrow \infty$.
In practice we of course use a non-vanishing value of $\Delta z$ and 
a finite value of $N_{\rm test}$. In the following we show how the
numerical solutions converge when $\Delta z$ is decreased and $N_{\rm test}$
is increased.

\begin{figure}[th]
\includegraphics[width=8.6cm]{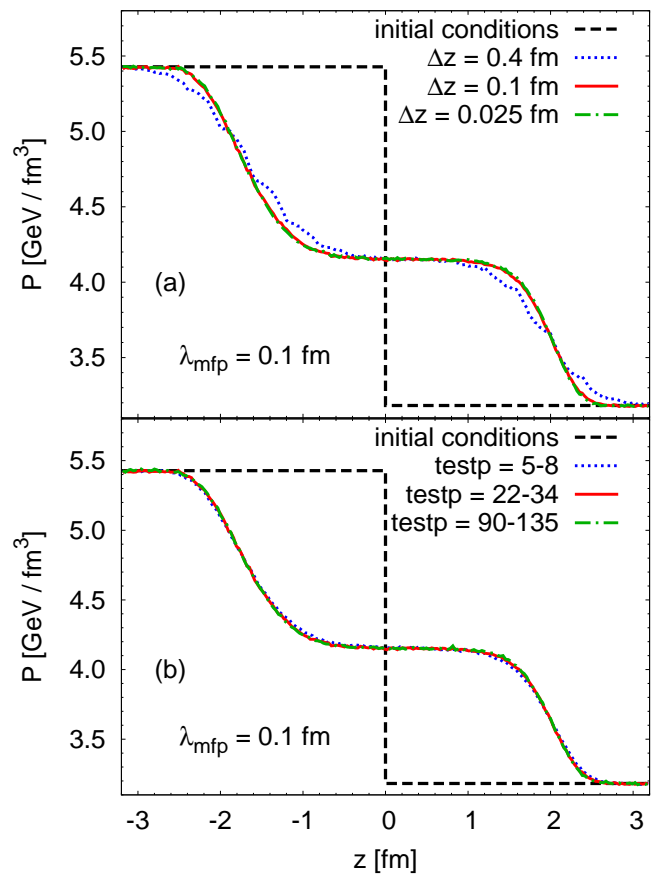}
\caption{(Color online) Cell size (a) and number of test particle
(b) dependence in the BAMPS simulation at $t = 3.2$ fm/$c$. The initial
conditions are chosen as $T_0 = 0.4$ GeV and $T_4 = 0.35$ GeV.
The mean-free path is $\lambda_{\rm mfp} = 0.1$ fm.
In (a) the pressure profile is shown for different cell size
$\Delta z = 0.4, 0.1, 0.025$ fm and constant $N_{\rm test}$.
In (b) we use a fixed $\Delta z = 0.1$ fm and
different numbers of test particles (testp) per cell. 
}
\label{fig:cell}
\end{figure}

The initial condition for this study is $T_0 = 0.4$ GeV and $T_4 = 0.35$ GeV.
Figure~\ref{fig:cell} shows the pressure profile at a time $t = 3.2$ fm/$c$
for a constant mean-free path $\lambda_{\rm mfp} = 0.1$ fm.
Results in the upper panel are obtained by varying $\Delta z$ and keeping
$N_{\rm test}$ unchanged. The number of
test particles in the cells is between 22 and 34, depending on the local temperature.
We see that convergence is reached when $\Delta z = \lambda_{\rm mfp}$.
Further decrease of $\Delta z$ does not lead to noticeable changes.
The lower panel of Fig.~\ref{fig:cell} shows the results for a fixed
$\Delta z$ and varying number of test particles per cell. Here we do not see significant
changes even for a small test-particle number in the cells. However, as we will
show in the next subsection, a small number of testparticles in cells causes
large fluctuations in each event, and thus affects heat flow, for instance.

Results from BAMPS, which will be presented in the rest of the paper,
are obtained by setting $\Delta z \le \lambda_{\rm mfp}$ and taking at least 15
test particles in each cell. We note that for these calculations a different
initial condition, $T_0 = 0.4$ GeV and $T_4 = 0.2$ GeV, is chosen. For
this case a faster convergence has been observed.

\subsection{Numerical solutions of the relativistic Riemann problem 
near the perfect-fluid limit}
\label{subsec:ideal_riemann}

The Riemann problem, which is analytically solvable 
in the perfect-fluid limit, is an important test case
for both the kinetic and the fluid-dynamical calculations.
In this section we show that both approaches can reproduce
the analytic solution very accurately, and also discuss
possible numerical uncertainties.

In BAMPS we cannot exactly reach the perfect-fluid limit,
but we can choose a very small physical viscosity
$\eta = 0.001 s$ to simulate an ideal fluid numerically. 
Use of even smaller viscosities, or equivalently larger cross sections, 
would require a better resolution (smaller $\Delta z$ and larger 
$N_{\rm test}$), which is computationally very time consuming.

On the other hand, for vSHASTA we can choose $\eta = \kappa_q = 0$, which solves
the relativistic Euler equations instead of the IS equations.
However, as explained earlier, because of the approximative nature of 
the numerical algorithm, we always have some residual numerical 
viscosity in the calculations.

\begin{figure}[th]
\includegraphics[width=8.6cm]{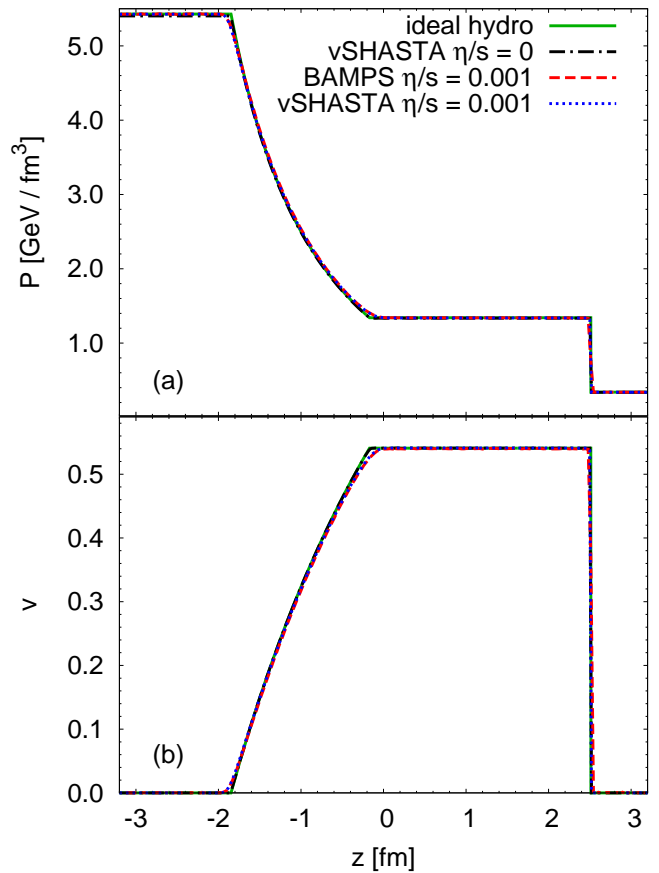}
\caption{(Color online) The analytic and numerical solutions of the
relativistic Riemann problem for the pressure (a) and velocity (b).
The initial conditions at $t = 0$ are
$T_0 = 0.4$ GeV and $T_4 = 0.2$ GeV.
The full lines are the analytic solutions at $t = 3.2$ fm/$c$.  
The results from BAMPS for $\eta/s = 0.001$ are shown
with dashed lines, the vSHASTA results in the perfect-fluid limit
with dash-dotted lines, and those for $\eta/s = 0.001$ with dotted lines.
}
\label{fig:pv_perfect}
\end{figure}

\begin{figure}[th]
\includegraphics[width=8.6cm]{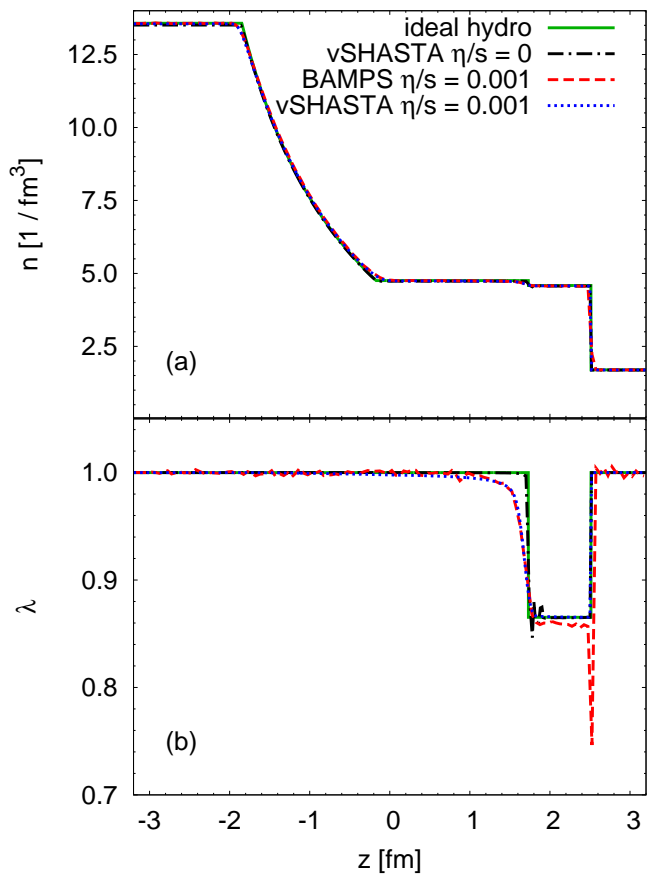}
\caption{(Color online) Same as in Fig.\ \ref{fig:pv_perfect}. 
The upper panel (a) shows the particle density while the lower panel (b) 
shows the fugacity.}
\label{fig:nl_perfect}
\end{figure}

In both calculations the initial state was chosen such that 
$T_0 = 0.4$ GeV and $T_4 = 0.2$ GeV. 
Figure~\ref{fig:pv_perfect} shows the pressure $p$
and velocity $v$, while Fig.~\ref{fig:nl_perfect}
shows the LRF particle density $n$ and fugacity $\lambda$
profiles at time $t = 3.2$ fm/$c$. 

The BAMPS results for $p$, $v$, and $n$ agree well with those of the
ideal fluid-dynamical solution, except for small deviations around the
discontinuities separating different regions. These deviations
are expected to appear because of the small but non-vanishing 
physical viscosity used in BAMPS calculations,
and are best seen in the fugacity profile.

Nevertheless, the BAMPS results also deviate from vSHASTA using
the same $\eta/s$ value; this is especially visible
in the fugacity profile between the contact
discontinuity and the shock front, where one observes a peak in the BAMPS result.
We will demonstrate later that this deviation is caused by numerical
fluctuations.

\begin{figure}[th]
\includegraphics[width=8.6cm]{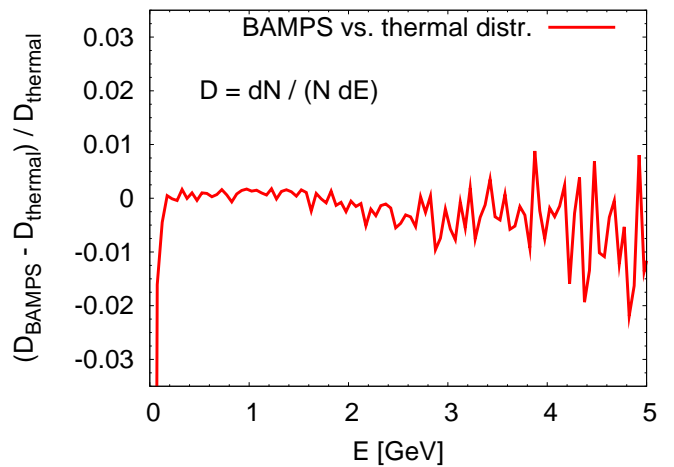}
\caption{(Color online) Relative difference of energy distribution
of particles at the 
plateau extracted from BAMPS and a local thermal equilibrium
distribution, Eq.\ \eqref{eq:thermalDistribution}.}
\label{fig:distribution_plateau}
\end{figure}

We also expect that particles in regions around discontinuities are
out of thermal equilibrium in the BAMPS calculations. However, this is
a small fraction of the whole system. The rest reaches and
approximately maintains thermal
equilibrium. In order to demonstrate this, we calculate the energy
distribution $dN / ( N \, dE )$ of particles in the region of the shock
plateau and compare it to the thermal one, which is
obtained by using Eq.\ \eqref{eq_distribution} with $a=0$,
\begin{equation} \label{eq:thermalDistribution}
\frac{dN}{N dE}=\frac{ e^{- \gamma E / T} \sinh(v \gamma E / T) E  }{2
T^2 \gamma^2 v}\,.
\end{equation}
Results are shown in Fig.~\ref{fig:distribution_plateau}, where we see
agreement within a few percent.

\begin{figure}[th]
\includegraphics[width=8.6cm]{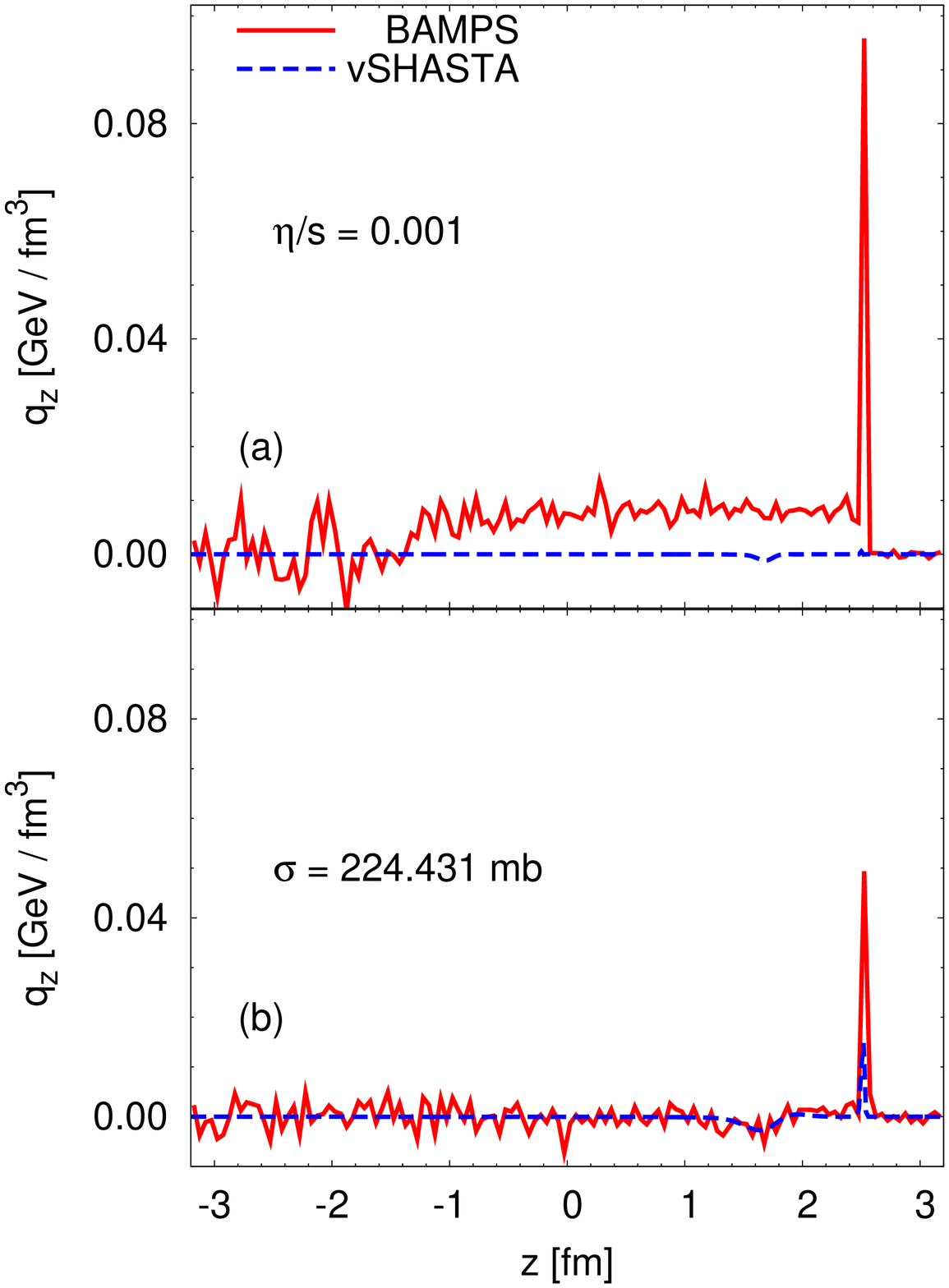}
\caption{(Color online) The heat-flow component (a) with fixed $\eta/s = 0.001$ 
and (b) with constant cross section $\sigma = 224.431\, \rm{mb}$.
The value of $\eta/s$ in the constant cross section simulation 
varies from 0.002 to 0.008.}
\label{fig:ideal_q_0.2}
\end{figure}

If the system is exactly in thermal equilibrium as described by the
ideal fluid-dynamical solution, the dissipative quantities, such as
the heat flow $q^z$, should vanish.
Figure~\ref{fig:ideal_q_0.2}(a) shows the heat-flow profiles
from both vSHASTA and BAMPS calculations with the same initial 
condition with $\eta/s=0.001$. Heat flow 
in vSHASTA is practically zero, while in BAMPS it has a small
positive value between the rarefaction fan and the shock front.
This deviation between BAMPS and vSHASTA results
was already noticeable in the
fugacity profile, but not in the other quantities shown above.
At the shock front we see a peak, similar as for the
fugacity profile.

The deviation of $q^z$ from zero observed in 
Fig.~\ref{fig:ideal_q_0.2} (a) (except for the peak) 
seems to be a numerical artifact, because this disappears when a constant
cross section is used instead of a constant $\eta/s$ value, as shown
in Fig.~\ref{fig:ideal_q_0.2} (b). Here, the cross section is set to be
$\sigma = 224.431$ mb, which corresponds to $\eta/s = 0.002$ in the medium
with the higher initial temperature.
In Fig.~\ref{fig:ideal_q_0.2} (b) we see perfect agreement between BAMPS
and vSHASTA results, especially comparing them at the small drop at 
$z \approx 1.6$ fm.
The peak at the shock front becomes smaller, although it is larger than
the peak from the vSHASTA calculation.

\begin{figure}[th]
\includegraphics[width=8.6cm]{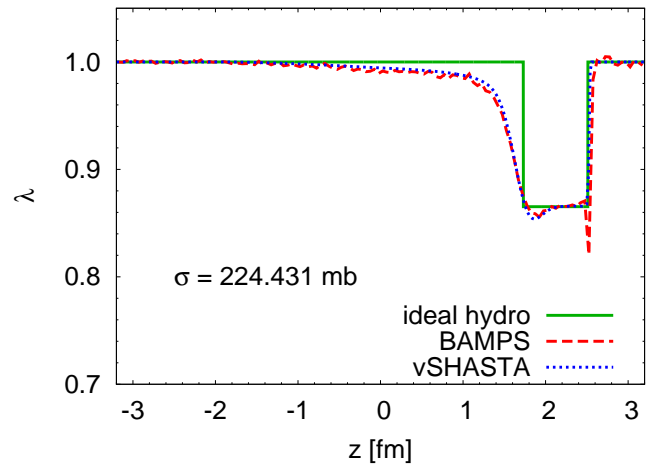}
\caption{(Color online) Fugacity calculated using a constant cross section
$\sigma = 224.431\, \rm{mb}$.}
\label{fig:fuga}
\end{figure}

Also, the difference in the fugacity between BAMPS and vSHASTA with
the same $\eta/s$ value, as observed in the lower panel of 
Fig.~\ref{fig:nl_perfect}, disappears almost completely when using
a constant cross section, as seen in Fig.~\ref{fig:fuga}.

On the shock plateau where the flow velocity and the LRF particle and 
energy density ($v$, $n$, $e$) are constant, one expects that the cross 
section is also constant for a constant $\eta/s$. However,
in a single event, thermodynamic quantities fluctuate, such that one
would use a smaller cross section (larger shear viscosity) 
in a cell with larger energy
(entropy) density in order to keep $\eta/s$ constant, and a larger
cross section in a cell with smaller energy density.
Therefore, although the results shown in
Figs.~ \ref{fig:nl_perfect} (lower
panel) and \ref{fig:ideal_q_0.2} (a) are averaged over 1000 events,
the deviations between BAMPS and vSHASTA results are likely to come
from the fluctuations in single BAMPS events. These can be reduced
by performing simulations with a much larger $N_{\rm test}$.
We have confirmed that in this case the difference between BAMPS and
vSHASTA solutions decreases.

\section{Viscous solutions of the relativistic Riemann problem}
\label{sec:viscous_riemann}

In this section we study the relativistic Riemann problem
at different non-zero viscosities. 
We will show that for small viscosities both the fluid-dynamical and kinetic 
approaches are in good agreement, especially at late times. 
However, with increasing viscosity this agreement fades and ultimately breaks
down when the fluid-dynamical description leads to results which are inconsistent 
with kinetic calculations.
Thus the main motivation of this study is to find the conditions of 
this break-down and then quantify the reach and limits of 
the dissipative fluid-dynamical description.

We note that all results shown below are calculated using
the Landau frame. In these test cases heat flow is small; therefore 
the differences between the Landau or Eckart frames are very small, even
for large values of $\eta/s$.

Kinetic theory can correctly treat the Riemann problem from
the nearly perfect limit to the free-streaming limit.
This has been previously demonstrated in 
Refs.~\cite{Bouras:2009nn,Bouras:2009zz} using BAMPS.
Another promising method to investigate the Riemann problem is based
on the lattice Boltzmann approach and has been recently reported in
Ref.~\cite{Mendoza:2010zz}. In contrast to kinetic theory, the validity
of IS theory requires that
the system stays close to local thermal equilibrium and the Knudsen
number $Kn$ is small during the whole evolution.

In the special case of the Riemann problem, at early times of the evolution,
the local Knudsen number is large where the density gradients are large,
even if the viscosity is small. Also the system around the discontinuity 
is far from equilibrium. Then in this region the IS theory of dissipative
fluid dynamics is expected to fail to describe the evolution correctly.

However, because of the viscosity and heat conductivity the gradients will be smoothed out 
later, hence providing better conditions for the IS fluid-dynamical description.
How close the solution will be to the kinetic one depends on the value of the
Knudsen number as demonstrated later in Sec.~\ref{global_Knudsen}.

In the next subsections we will show results at fixed times but for different
values of $\eta/s$. However, solutions at time $t$ with 
shear viscous coefficient $\eta$ correspond to solutions at 
time $a\,t$ with shear viscous coefficient $a\, \eta$, where
$a$ is some arbitrary constant. 
This scaling behavior is discussed later in Sec.~\ref{sec:scaling_behaviour}.

\subsection{Small viscosity}

\begin{figure}[th]
\includegraphics[width=8.6cm]{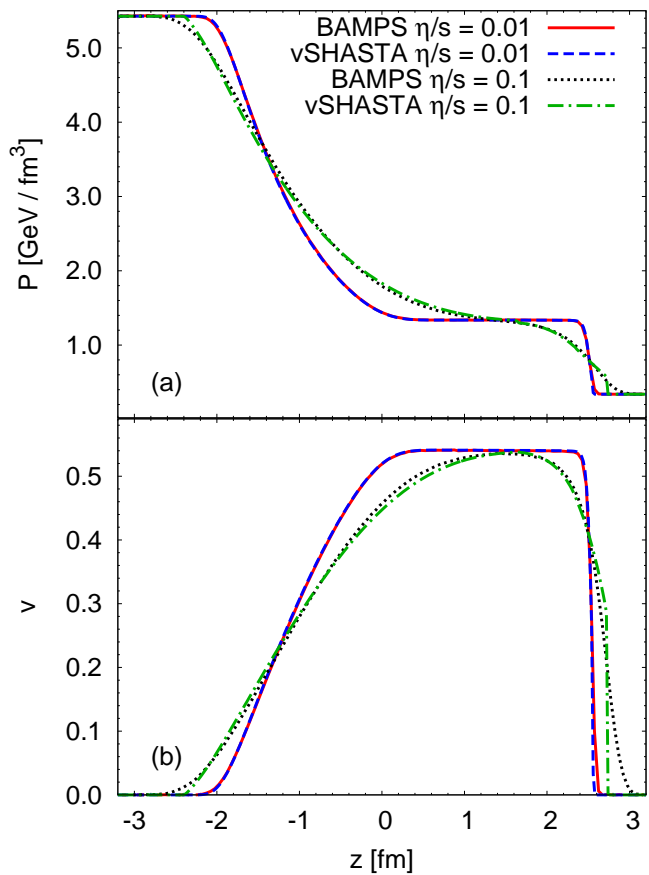}
\caption{(Color online) The same as Fig.\
\ref{fig:pv_perfect}, for $\eta/s = 0.01$ and 0.1.}
\label{fig:pv_viscous}
\end{figure}

\begin{figure}[th]
\includegraphics[width=8.6cm]{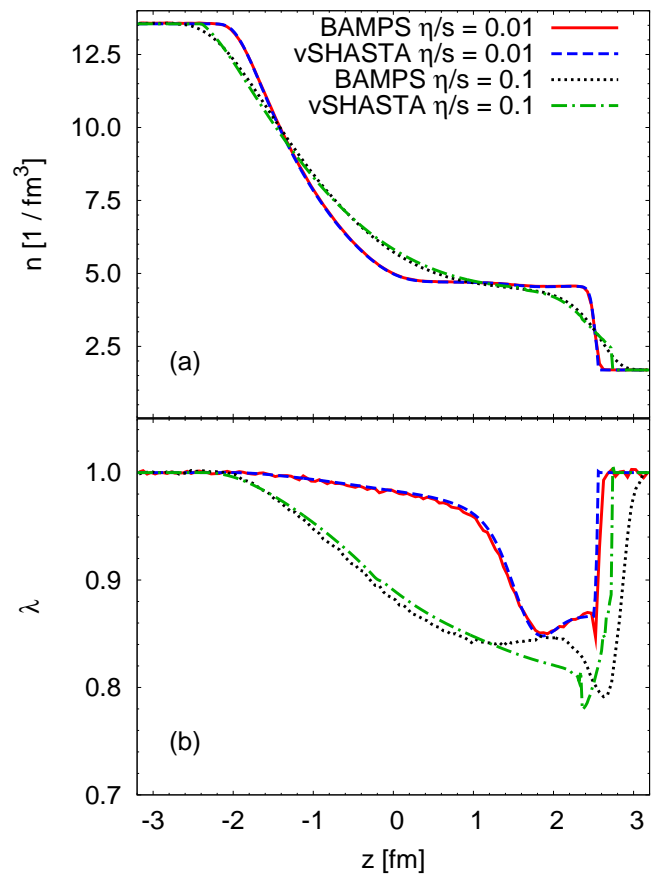}
\caption{(Color online) The same as Fig.~\ref{fig:pv_viscous}, for
(a) particle density and (b) fugacity.}
\label{fig:nf_viscous}
\end{figure}

\begin{figure}[th]
\includegraphics[width=8.6cm]{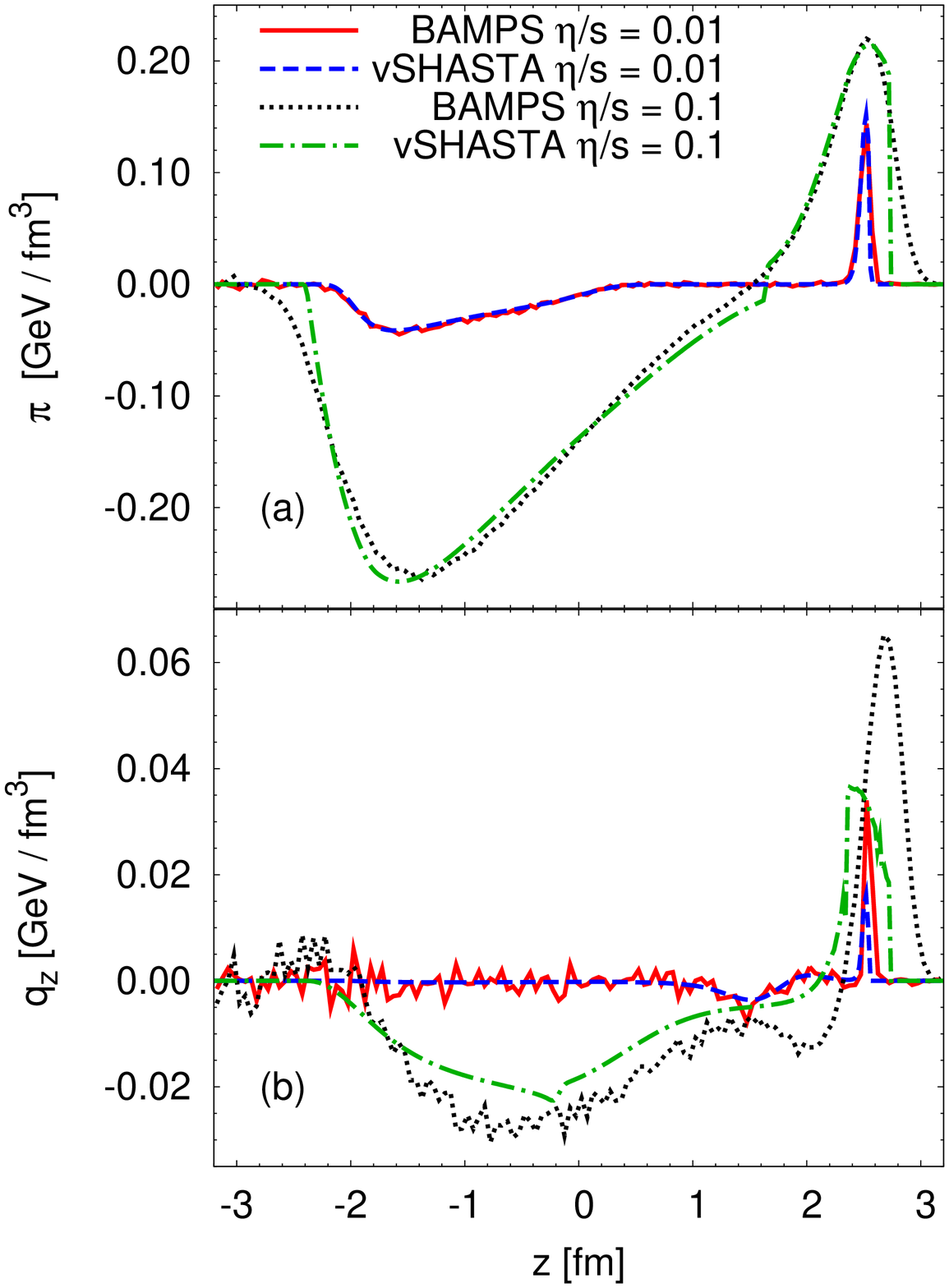}
\caption{(Color online) The same as Fig.~\ref{fig:pv_viscous}, for
(a) shear pressure  and (b) heat flow.}
\label{fig:qpi_viscous}
\end{figure}

We use the same initial conditions as given in the previous section,
but now for two different values of the shear viscosity to 
entropy density ratio,  $\eta/s = 0.01$ and $\eta/s = 0.1$. 
Figure~\ref{fig:pv_viscous} shows the pressure $p$ and velocity $v$,
Fig.~\ref{fig:nf_viscous} shows the LRF particle density $n$
and the fugacity $\lambda$, and Fig.~\ref{fig:qpi_viscous}
shows the profiles of the shear pressure $\pi$ and the heat flux $q^z$
at time $t = 3.2$ fm/$c$ from both BAMPS and vSHASTA calculations.

In the dissipative case the characteristic structures of the 
perfect-fluid solution can still be found in the late stages of the evolution, 
since it takes a finite time for the structures to form; see Sec.~\ref{sec:scaling_behaviour}.
However, instead of a discontinuous shock front, a contact discontinuity, and 
sharp rarefaction tails, we get continuously changing profiles, that is, 
dissipation leads to the smoothing and broadening of these 
characteristic structures.

Further differences compared to the perfect-fluid case are that 
the head and the tail of the rarefaction fan 
and the shock front propagate faster into the undisturbed matter.
However, for the shock wave this happens only until the shock plateau 
is formed. After that the velocity of the shock wave is the same as
for the perfect-fluid case. Similarly, the velocity of the plateau
does not change from the perfect-fluid solution.

For the smaller shear viscosity to entropy density ratio, $\eta/s = 0.01$,
the agreement between
BAMPS and vSHASTA results for all macroscopic quantities is excellent
within statistical fluctuations, although any definite conclusions
regarding the heat fluxes are hard to draw because of large fluctuations. 
The shock front from both calculations is also in very good agreement. 
However, a closer inspection reveals that vSHASTA gives slightly 
steeper profiles than BAMPS.

Increasing the viscosity to $\eta/s = 0.1$ leads to noticeable 
differences between the approaches.
The most pronounced difference can be seen in the shock front: 
vSHASTA provides a too sharp profile at the right edge of the front,
while in the BAMPS calculation the matter is diffused faster in the 
low-density region. 
This can also be seen in the rarefaction fan where the kink at the 
left edge survives. 
Another difference can be seen by inspecting the fugacity and
shear pressure 
in the region where the contact discontinuity would be in the perfect-fluid case. 
In this region the vSHASTA calculation returns an overall smaller fugacity, 
and also a sharp kink in the shear pressure profile. 
These differences will be enhanced for larger viscosities as seen in the
next subsection.

\subsection{Large viscosity}

\begin{figure}[th]
\includegraphics[width=8.6cm]{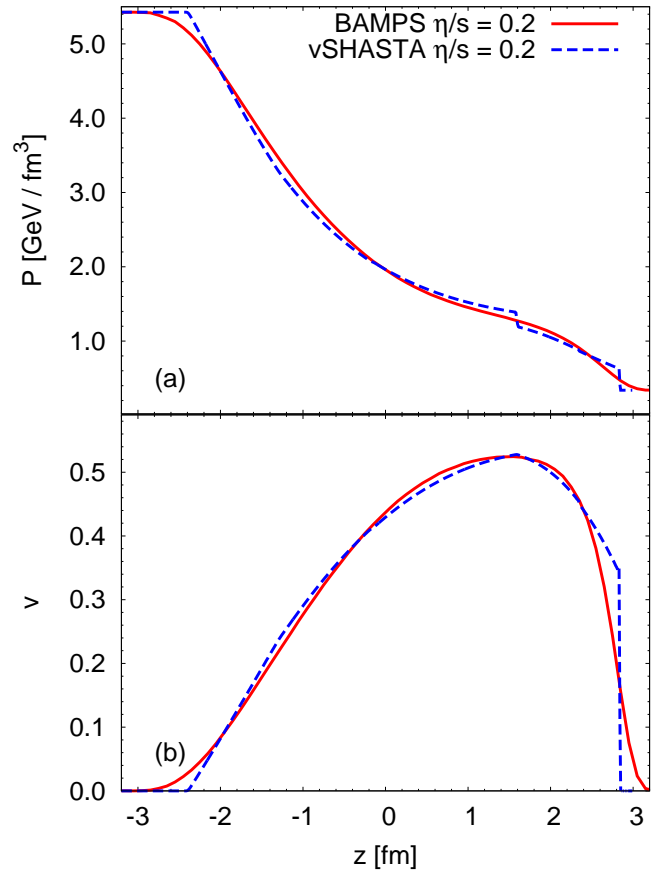}
\caption{(Color online) The same as Fig.~\ref{fig:pv_viscous}, 
for $\eta/s = 0.2$.}
\label{fig:pv_break}
\end{figure}

If the value of $\eta/s$ is further increased, that is, $\eta/s = 0.2$, 
we start to see much larger deviations between vSHASTA and BAMPS results.
These are presented in Fig.~\ref{fig:pv_break} for the pressure 
and velocity profiles. 

The most salient difference is seen in the pressure
profile: in the vSHASTA calculation a part of the initial 
discontinuity survives near the contact discontinuity, 
at $z \sim 1.5$ fm even after $t=3.2$ fm/$c$. 
In contrast, this kind of structure is not seen in the BAMPS calculation. 
A similar shock structure was also observed in Ref.~\cite{Denicol:2008rj}, 
called the "double-shock" phenomenon by the authors.
It is important to note that in that case a different numerical method,
namely, the smoothed particle hydrodynamics, was used to solve the
equations of dissipative fluid dynamics, 
corresponding to the simplified IS equations without heat conductivity. 
The simplified or truncated IS equations only take into account the relaxation 
term to describe the time evolution
of dissipative quantities. 

In the fluid-dynamical calculation this discontinuity 
originates from the initial discontinuity. 
In the early stage of the evolution the effective 
pressure $p+\pi$ and velocity take almost constant 
values near the discontinuity. 
The velocity and the gradient of the effective
pressure are the driving forces of the expansion. 
Therefore, if they are constant, nothing
happens to the structures in the solution and 
the original discontinuity disappears very slowly, 
such that parts of it are still visible in the later stages 
of the evolution.

A similar difference between the BAMPS and vSHASTA results
is seen at the right
edge of the shock wave. This part of the profile is again not
described correctly by IS fluid dynamics, and the difference is 
already visible for smaller values of $\eta/s$. 
The same is true also for the head of the rarefaction fan, 
although it is less
visible for small viscosities. 
This kind of discontinuous behavior can be seen in all relevant 
fluid-dynamical quantities.

In the BAMPS calculation the original discontinuity 
disappears immediately. 
This is because in kinetic theory the evolution near the 
very steep density gradient is well approximated by free streaming 
or diffusion of particles, which smooths out all sharp structures 
very rapidly.
Free streaming of particles drives the system immediately far out of thermal
equilibrium, hence cannot be correctly described by second-order
dissipative fluid dynamics. 

This phenomena was studied and explained in non-relativistic systems 
M. Torrilhon $et$ $al.$\cite{Torrilhon}.
They concluded that the viscous fluid-dynamical solutions of the Riemann problem 
actually lead to discontinuous solutions.
For example, in the non-relativistic 13-field equations the system has five
instead of three characteristic waves.
Although dissipation leads to the attenuation of these waves and smoothing of discontinuities, 
this can only happen after a sufficiently long time.
In case we include higher moments of the distribution function we will find
more characteristics and therefore more discontinuities but with smaller amplitude.
There is an infinite number of moments, which form a hierarchy of equations.
Therefore by taking into account higher moments the approximations to 
the Boltzmann equation become more precise, which in turn leads to a better 
approximation for smooth profiles.
On this account we note that there are more recent studies showing that with a special 
regularization technique the Grad's 13-moment method leads to much better 
results \cite{Struchtrup}.
So far these methods have been studied only in the non-relativistic case, 
but nevertheless they point toward a solution.

\subsection{Heat-flow problem}

\begin{figure}[th]
\includegraphics[width=8.6cm]{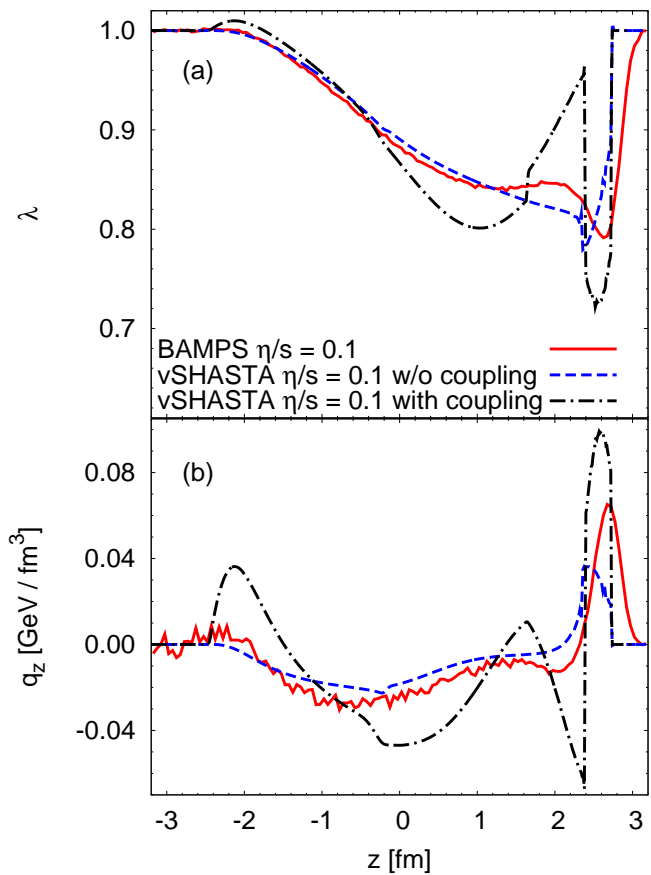}
\caption{(Color online) The fugacity (a) and heat flow (b) profiles with and
without coupling to the shear pressure in the heat-flow equation,
$I_{q3}^z$ in Eq.\ \eqref{relaxationEq_heatAndShear} }
\label{fig:fq_coupling}
\end{figure}

As already mentioned in Sec.~\ref{subsec:IS_1plus1} we neglected
the term that couples the heat flux to the shear pressure from
the heat-flow equation \eqref{relaxationEq_heatAndShear}. 
The reason is that, if this term is included,
the good agreement of heat flow and fugacity between
BAMPS and vSHASTA is lost even for small viscosity. 
This is demonstrated in Fig.~\ref{fig:fq_coupling}, 
where we show the fugacity $\lambda$ and heat flux $q^z$
for $\eta/s = 0.1$ with and without this coupling term.
The profiles change completely, and there is no support 
from BAMPS for structures induced by the coupling. 
For other quantities this coupling term has a very small effect.

The reason that this single term can become dominant
is that in the viscous Riemann problem, the heat flow is typically
one order of magnitude smaller than the shear pressure. Thus,
the coupling term in the heat equation can be large when the
shear pressure is large, even if it is formally only 
a second-order correction.

\subsection{Global Knudsen number analysis}
\label{global_Knudsen}

In order to better quantify and measure the applicability of IS theory,
we define the relative difference between the BAMPS and vSHASTA calculations as
\be\label{eq:rel_diff}
\left<\frac{\delta e}{e}\right>^2 = \frac{1}{\Delta z}\int dz \left(\frac{\delta e}{e}\right)^2,
\ee
where $\delta e$ is the difference in energy density between the BAMPS 
and the vSHASTA calculations. We recall that $e = 3p$ for a massless gas.
The integral is evaluated from the head of 
the rarefaction fan to the tail of the shock wave; hence the constant 
temperature regions to the left and right are not included. The width of this 
region is denoted as $\Delta z$.
Similarly, the average macroscopic length scale can be estimated from the 
average energy density gradient as
\be
\label{eq:ave_L}
L_e^{-1} = \left|\frac{1}{\Delta z}\int \frac{1}{e}\frac{\partial e}{\partial z} dz\right| 
= \frac{1}{\Delta z} \ln{\frac{e_0}{e_4}} \, ,
\ee
where $e_0$ and $e_4$ are the initial energy densities on the left- and
right-hand sides of the initial discontinuity. 
Hence, an average Knudsen number relevant for this study can be defined as
\be
\label{eq:ave_Kn}
Kn_e = \frac{\lambda_{\rm mfp}}{L_e} \, ,
\ee
where $\lambda_{\rm mfp}$ is the mean-free path in the low-temperature region,
that is, it is the largest mean-free path.
This definition smooths the rapid changes compared to a local Knudsen number
and makes the comparisons between calculations feasible. 
Note that the Knudsen number \eqref{eq:ave_Kn} is similar to that
introduced in Ref.~\cite{Bouras:2009nn}.

Since $\lambda_{\rm mfp} \sim \eta/(Ts)$ and $L_e \sim t$, where $t$
is a given time during the evolution, we also have
\begin{equation} \label{Knudsenscaling}
Kn_e \sim \frac{\eta}{s}\, \frac{1}{T t}\;.
\end{equation}
Therefore, for a given temperature, $Kn_e$ stays constant
if we scale $\eta/s$ and $t$ by the same factor.

\begin{figure}[th]
\includegraphics[width=8.6cm]{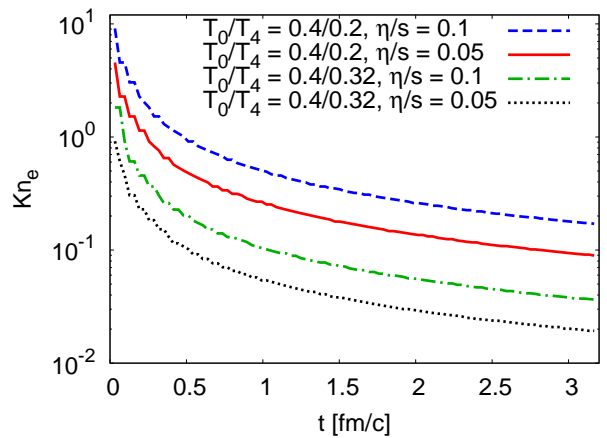}
\caption{(Color online) Time evolution of the average Knudsen number, $Kn_e$,
for different initial temperature ratios and different values of $\eta/s$.}
\label{fig:global_kn}
\end{figure}

\begin{figure}[th]
\includegraphics[width=8.6cm]{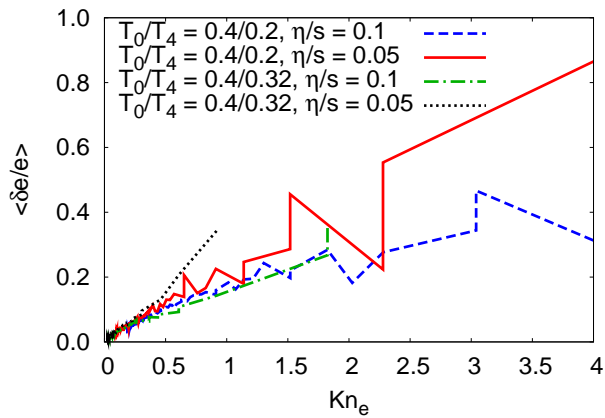}
\caption{(Color online) The relative difference of kinetic and fluid-dynamical
calculations for different initial temperature ratios and different values of $\eta/s$ 
as a function of the average Knudsen number.}
\label{fig:global_kn_err}
\end{figure}

Figure~\ref{fig:global_kn} shows the time evolution of the average
Knudsen number for two different initial conditions and two different viscosities. 
In all cases the Knudsen number is initially large but decreases rapidly as the
system expands. 
This happens as a function of the initial temperature difference and viscosity.

On the other hand, the relative difference between BAMPS and vSHASTA calculations 
decreases with decreasing average Knudsen number.
This is shown in Fig.~\ref{fig:global_kn_err} where we can see that for small Knudsen 
numbers the different solutions converge to approximately one curve.
This also means that to good approximation the Knudsen number alone determines the 
applicability of IS theory. We read off Fig.\ \ref{fig:global_kn_err} that the
differences between BAMPS and vSHASTA are less than $10 \%$ for $Kn_e < 1/2$.

\section{Formation of shock waves and its scaling behaviour}
\label{sec:scaling_behaviour}

In ideal fluid dynamics shock waves are formed immediately after removing
the membrane that separates matter with different temperatures. This happens
because the Knudsen number $Kn_e$ vanishes at any time. For non-vanishing
viscosity $Kn_e$ is large at early times and, thus, the formation of 
shock waves occurs later, when $Kn_e$ becomes smaller, for example, in the
case for a constant $\eta/s$ value, as demonstrated in
Fig.~\ref{fig:global_kn}.

\begin{figure}[th]
\includegraphics[width=8.6cm]{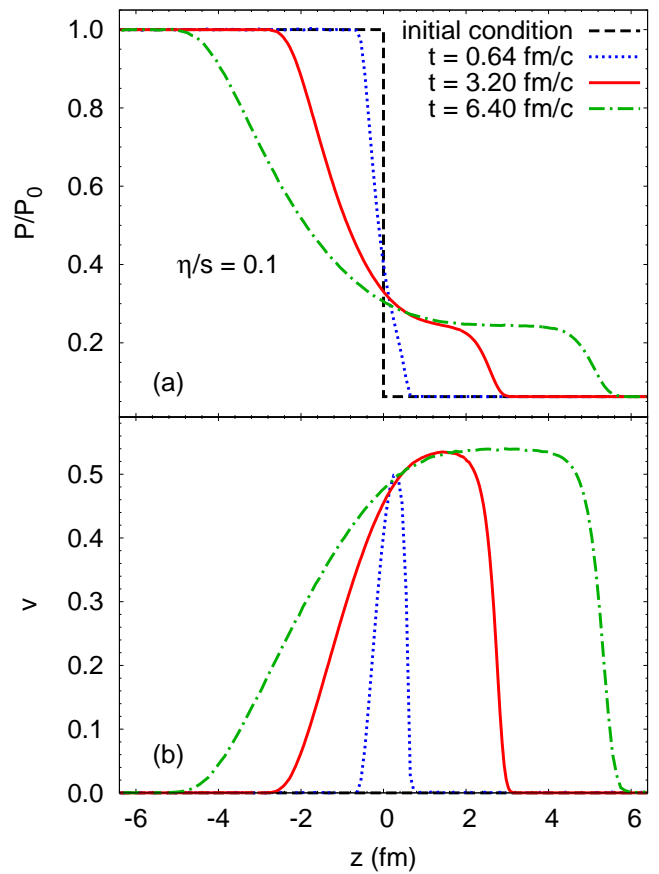}
\caption{(Color online) The time evolution of the shock-tube problem 
for $\eta/s = 0.1$. The initial condition is the same as in
Fig.\ \ref{fig:pv_perfect}.}
\label{fig_evolution}
\end{figure}

Figure~\ref{fig_evolution} shows the pressure and velocity profiles at 
various times for $\eta/s = 0.1$. At the early time $t = 0.64$ fm/$c$
shock waves have not yet developed. The pressure profile looks like that in
the strong diffusive case of free-streaming particles.
At $t = 6.4$ fm/$c$ we observe a characteristic shock plateau that
clearly separates the shock front from the rarefaction wave,
as in the ideal-fluid case. The intermediate time $t = 3.2$ fm/$c$ is
the time scale at which the shock plateau is being formed and 
the maximum of the velocity distribution $v(z)$ reaches the value 
$v_{\rm plat}$ of the ideal-fluid solution.
We define this time scale as the formation time of shocks.

\begin{figure}[th]
\includegraphics[width=8.6cm]{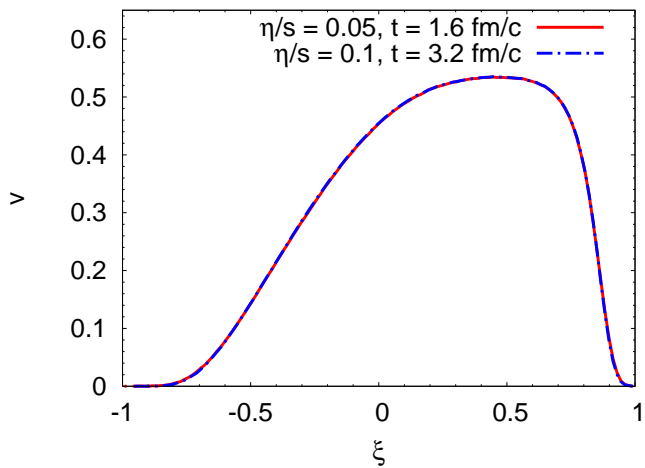}
\caption{(Color online) The scaling behavior for the shock-tube problem.
The velocity profiles are shown as a function of the similarity variable $\xi = z/t$
for $\eta/s=0.1$ and $\eta/s = 0.05$ at times $t = 3.2$ fm/$c$ 
and $t = 1.6$ fm/$c$, respectively.}
\label{fig_scaling}
\end{figure}

The only intrinsic length scale in the microscopic approach is
the particle mean-free path. Therefore, if we rescale the mean-free
path by a constant factor of $a$, we expect the time scale
for the evolution of matter to change accordingly. Since
$\eta/ s \sim \lambda_{\rm mfp}$, we expect that profiles
calculated at time $t$ for a given value of $\eta/s$ agree with
those at a time $a\, t$ for a viscosity-to-entropy density
ratio $a\, \eta/s$.
This is demonstrated in Fig.~\ref{fig_scaling} where we show the
velocity profiles as a function of the similarity variable
$\xi = z/t$. 

The pressure and the velocity profile as a function of $\xi$ are
determined by the Knudsen number $Kn_e$ \eqref{eq:ave_Kn}. According
to Eq.\ \eqref{Knudsenscaling},
$Kn_e$ is the same for the calculation with $\eta/s=0.1$ at $t=3.2$ fm/$c$ and
that with $\eta/s=0.05$ at $t=1.6$ fm/$c$. Therefore, 
the pressure $p(z,t;\eta/s)/p_0$ and velocity
$v(z,t;\eta/s)$ are functions only of $\xi$ and $Kn_e$,
that is, $p(z,t;\eta/s)/p_0=F(\xi; Kn_e)$, and similarly for the
velocity. For decreasing $Kn_e$ the plateau
of the velocity profile in Fig.~\ref{fig_scaling} will be growing and
approaches the shape
of the ideal-fluid case shown in the lower panel of Fig.~\ref{fig:riemann_nv}.

This scaling behavior holds only for
initial conditions with
a discontinuity in pressure. If the discontinuity is changed to a smooth 
transition, the non-zero width $\Gamma_{\rm tr}$ of the 
transition region introduces another length scale. 
Then, as a function of the similarity variable, this transition
region would change under a rescaling $t \rightarrow a t$,
$\xi_{\rm tr} = \Gamma_{\rm tr}/t \rightarrow \Gamma_{\rm tr}/(at)$, 
and thus cause a different gradient in the transition region as a function of $\xi$. 
Because of the different initial situations,
evolutions in $\xi$ for the same $Kn_e$ are not identical.

\section{Conclusions}
\label{sec:conclusions}

In this work we have studied the formation and evolution of relativistic shock 
waves in dissipative matter with non-zero shear viscosity 
and heat conductivity by solving the relativistic Riemann problem.
This was accomplished by using both relativistic kinetic theory 
and relativistic dissipative fluid dynamics.
The relativistic kinetic approach solves the Boltzmann equation
by using the BAMPS code \cite{Xu:2004mz}. 
The fluid-dynamical approach is based on Israel and 
Stewart theory \cite{Israel:1979wp} and was solved numerically 
by the vSHASTA method for hyperbolic equations.

After extensive comparisons between the two approaches, we found 
the following: When the viscosity is zero, both give 
equivalent results. 
It was demonstrated that both approaches reproduce the 
analytic solutions of the Riemann problem in the perfect-fluid limit 
and the numerical results converge when the numerical 
resolution is sufficiently high.

Departing from the perfect-fluid limit, for cases when the viscosity is small, 
the agreement between kinetic theory and IS theory is still excellent. 
As the viscosity increases the agreement between the approaches starts 
to deteriorate.
For even larger values of the $\eta/s$ ratio, IS theory develops 
discontinuities which survive even after long times.
These sub-structures are not supported by the kinetic simulations.
They are an artifact of the method of moments \cite{Torrilhon} on
which 
IS theory is based.
However, we also argued that part of this discrepancy can be understood 
to result from the inapplicability of IS theory for large 
Knudsen numbers.

Quantitative statements about the applicability 
of IS theory are difficult to make in the current setup, mainly 
because the early evolution is not well described by fluid dynamics.
This also affects the late evolution of the system, and therefore it is
difficult to clearly separate between the artifacts from the early 
fluid-dynamical solutions and the effects of large Knudsen number.

However, we showed that a quantitative analysis in terms
of an average Knudsen number is possible in such a way that 
it gives a good measure for the applicability of the IS theory.
In accordance with Ref.~\cite{Huovinen:2008te} we found that for 
$Kn_e<1/2$ the difference between kinetic theory and IS theory
is less than $\sim 10$ \%.

The shear pressure profile was reasonably well described
by the IS equations from small to moderate viscosities.
Furthermore, we also found that the results are quite insensitive 
to the second-order terms in the IS equation for the 
shear viscosity. 
However, the same is not true for heat flow. 
At very small viscosities the heat flow was quite well described, including 
all terms in the IS equation.
At larger viscosity the results are very sensitive to the 
coupling term between the heat flow and shear viscosity.
In the Riemann problem studied here, 
the heat flow is an order of magnitude smaller than 
the shear pressure, and the coupling to shear dominates the 
behavior of the heat flow. 
This coupling gives a too large contribution in the heat-flow component,
which is not supported by the kinetic calculations.
Whether the inclusion of all second-order terms in the IS
equations will cure the heat-flow problem and improve the overall 
applicability  of the fluid-dynamical approach will be studied in 
the future.

\section*{Acknowledgements}

The authors are grateful to P.\ Huovinen, G.\ Denicol, B.\ Betz,
L.P.\ Csernai, J.A.\ Maruhn, and H.\ St\"ocker
for discussions 
and to the Center for Scientific 
Computing (CSC) at Frankfurt University for the computing resources.
I.\ B. is grateful to HGS-Hire.
E.\ M. acknowledges support by OTKA/NKTH 81655 and
the Alexander von Humboldt foundation. The work of H.\ N. was supported by
the Extreme Matter Institute (EMMI).

This work was supported by the Helmholtz International Center
for FAIR within the framework of the LOEWE program 
launched by the State of Hesse.



\end{document}